\documentclass{article}

\usepackage{amsmath,amssymb,amsthm}
\usepackage{url}
\usepackage{proof}
\usepackage{stmaryrd}
\usepackage{parskip}
\usepackage{fullpage}

\newcommand{\abs}[4]{{#1}\, #2\! : \! #3.\, #4}
\newcommand{\absu}[3]{{#1}\, #2.\, #3}

\mathchardef\mhyph="2D 
\begin{document}

\newcommand{\interp}[1]{\llbracket #1 \rrbracket} 
\newtheorem{theorem}{Theorem}
\newtheorem{lemma}[theorem]{Lemma}
\newtheorem{definition}[theorem]{Definition}
\newtheorem{corollary}[theorem]{Corollary}

\title{Syntax and Typing for Cedille Core}

\author{Aaron Stump}

\maketitle

\section{Motivation}

This note proposes a syntax for annotated terms and type-computation
rules for them, for a core version of Cedille, to which full Cedille
is easily translatable.  Additionally, Cedille Core may be a good target
for other higher-level extrinsic type theories.

The goal is to have a version of Cedille for which typing can be
implemented straightforwardly, in a small, trustworthy checker.
Cedille as we are implementing it for interactive development of
proofs has a lot of features that are not needed in a lower-level
format intended just for confirming typings.  The checker for core
Cedille does not need to produce span information for the benefit of
an IDE, as our full implementation of Cedille does.  Furthermore, it
is acceptable for core Cedille to require more annotations on terms,
to make type checking easier to implement.

There is one downside of the approach taken here: there will be
notably more calls to check definitional equality of types, and hence
checking time might be a bit slower.  This is because when checking
(for instance) an application of $f : A \to B$ to $a : A'$, we must
check that $A$ and $A'$ are definitionally equal.  In bidirectional
checking, we synthesize the type $A \to B$ for $f$ , and then check
the argument $a$ against $A$.  The latter check will generally
decompose $A$, but could be quite a bit cheaper than testing
definitional equality of $A$ with some other (possibly identical!)
type.

\subsection{A few wrinkles}

There are a few places where the system below differs from what we
have developed so far in papers and/or our implementation (which is
usually ahead of the papers):

\begin{enumerate}
  \item The system below uses Curry-style type-level
    $\lambda$-abstractions (so no type with the bound variable), while
    Cedille as we have developed it so far uses Church-style ones (a
    type for the bound variable is an essential part of the
    expression).  We believe we know how to support Curry-style
    type-level $\lambda$-abstractions now; we did not know how to
    give the semantics for them before.  

  \item The system below omits lifting, which is needed in our
    approach for large eliminations.  We have an approach to lifting
    in~\cite{stump17a}, but the way it is done there is somewhat
    complicated, and furthermore we believe that it is incomplete: there
    are some lifting terms (with free variables) that should be semantically
    equivalent but which are not definitionally equal.  So we are
    anticipating needing to rework lifting, which will likely not be
    an easy theoretical job (though probably not complicating the implementation
    of Core Cedille very much).  So I have left out lifting for now.

  \item Note that full Cedille currently does not allow $\forall$ at the kind level,
    and hence does not support erased $\lambda$-abstractions at the type level.  While
    we think we know enough now to support this feature semantically, we have not
    worked out the details yet and hence omit it.

    \end{enumerate}

\section{Syntax}
\label{sec:syntax}

Figure~\ref{fig:syntax} gives the syntax for terms of Cedille Core, in
the style of pure type systems~\cite{B92}: we have just one syntactic
category of terms, and we rely on the type system to distinguish
terms, types, and kinds (and a sole superkind $\Box$).  The constructs
are listed with some short comments to give a hint to the reader of
what they mean; the typing rules below should clarify the meaning
further.  Also, the constructs are listed in the figure in order from
highest precedence (most aggressively binding arguments) to lowest.
This ordering is a little different in a couple places from current
Cedille, but we have changed our parser recently and so may redo our
operator order to match this.  I believe this order is a little more
natural than the one we are currently implementing.  By the way, we
also plan to add support to our implementation of (full) Cedille for a
couple forms below that are currently not legal Cedille syntax
(dependent introductions, the particular form of $\rho$-terms).  In
the typing rules below, we will also use this definition:
\[
\textit{sorts}\ \mathcal{S} := \{ \star , \Box \}
\]
\noindent We will use $s$ as a metavariable ranging over $\mathcal{S}$.

\begin{figure}
\[
\begin{array}{llll}
  \textit{term variables}\ u & \ & \ &\ \\
  \textit{type variables}\ X & \ & \ &\ \\
  \textit{kind variables}\ k & \ & \ &\ \\
  \textit{variables}\ x & ::= & u\ |\ X\ |\ k &\ \\  
  \textit{pure terms}\ p & ::= & u\ |\ p\ p'\ |\ \absu{\lambda}{u}{p} &\ \\
  \textit{annotated terms}\ t & ::= & x\ & \textnormal{use of a variable} \\
  \ &\ & \star&\textnormal{kind for types}\\
  \ &\ & \Box&\textnormal{sole superkind} \\
  \ &\ & t.1 & \textnormal{project first view of a dependent intersection}\\
  \ &\ & t.2 & \textnormal{project second view of a dependent intersection}\\
  \ &\ & \beta\ t'\{t\} &\textnormal{proof of }t' \simeq t'\textnormal{, where the proof erases to }t  \\
  \ &\ & \delta\ T\ t &\textnormal{proves anything if}t\textnormal{ proves a certain false equation}  \\
  \ &\ & \varsigma\ t &\textnormal{symmetry of equality}  \\
  \ &\ & t\ t' & \textnormal{application of term to term}\\
  \ &\ & t\ \mhyph t' & \textnormal{application of a term to an erased argument}\\
  \ &\ & \rho\ t\ @\ x.t'\ \mhyph\ t''&\textnormal{equality elimination by type-guided rewriting}  \\ 
  \ &\ & \abs{\forall}{x}{t}{t'}\ &\textnormal{implicit product (quantify over erased argument)} \\
  \ &\ & \abs{\Pi}{x}{t}{t'}\ &\textnormal{explicit product (usual }\Pi\textnormal{-type)} \\
  \ &\ & \abs{\iota}{x}{T}{T'}\ &\textnormal{dependent intersection} \\
  \ &\ & \abs{\lambda}{x}{t}{t'} & \textnormal{usual }\lambda\textnormal{-abstraction} \\
  \ &\ & \abs{\Lambda}{x}{t}{t'} & \textnormal{erased }\lambda\textnormal{-abstraction}\\
  \ &\ & [ t , t' \ @\ x.t'' ] & \textnormal{introduce dependent intersection}\\
  \ &\ & \phi\ t\ \mhyph\ t'\ \{t''\} & \textnormal{when }t\textnormal{ proves }t'\textnormal{ and }t''
                                   \textnormal{ are equal, erase to }t'' \\
  \ & \ & [ x = t : t' ]\ \mhyph\ t'' & \textnormal{let }x\textnormal{ equal }t\textnormal{ of type }t'\textnormal{ in }t'' \\
  \ &\ & \{ p \simeq p' \} &\textnormal{equality between pure terms}
\end{array}
  \]
  \caption{Sytax for Cedille Core}
  \label{fig:syntax}
\end{figure}

\subsection{Syntactic check for being a term}
\label{sec:termcheck}

It is important that we only form equations between terms, because
Cedille's semantics does not support forming equations except where
both sides of the equation are terms.  Terms can be syntactically
distinguished from types and kinds as long as we use different
syntactic categories for term variables, type variables, and kind
variables.  That is why the syntax of Figure~\ref{fig:syntax}
distinguishes these.  The typing rule below for forming equations
requires the sides to be pure (i.e., unannotated) terms.

\section{Erasure}

When comparing terms for definitional equality, the rules below will
compare erased terms, using the erasure function defined in
Figure~\ref{fig:eraser}.

\begin{figure}
  \[
  \begin{array}{lll}
    |x| & = & x \\
    |\star| & = & \star \\
    |\Box| & = & \Box \\
    |t.1| & = & |t| \\
    |t.2| & = & |t| \\
    |\beta\ t'\{t\}| & = & |t|\\
    |\delta\ T\ t| & = & |t|\\
    |\varsigma\ t| & = & |t| \\
    |t\ t'| & = & |t|\ |t'| \\
    |t\ \mhyph t'| & = & |t| \\
    |\rho\ t\ @\ x.t'\ \mhyph\ t''| & = & |t''| \\
    |\abs{\forall}{x}{t}{t'}| & = & \abs{\forall}{x}{|t|}{|t'|}\\
    |\abs{\Pi}{x}{t}{t'}| & = & \abs{\Pi}{x}{|t|}{|t'|}\\
    |\abs{\iota}{x}{T}{T'}| & = & \abs{\iota}{x}{|t|}{|t'|} \\
    |\abs{\lambda}{u}{t}{t'}| & = &  \absu{\lambda}{u}{|t'|} \\
    |\abs{\lambda}{X}{t}{t'}| & = &  \abs{\lambda}{X}{|t|}{|t'|} \\
    |\abs{\Lambda}{x}{t}{t'}| & = &  |t'| \\ 
    |[ t , t' \ @\ x.t'' ]| & = & |t| \\ 
    |\phi\ t\ \mhyph\ t'\ \{t''\}| & = & |t''| \\
    |[ x = t : t' ]\ \mhyph\ t'' | & = & (\absu{\lambda}{x}{|t''|})\ |t|\\
    |\{ t \simeq t' \}|| & = & \{ |t| \simeq |t'| \}
  \end{array}
  \]
  \caption{Erasure for annotated terms}
  \label{fig:eraser}
\end{figure}

\section{Typing}

The type-checking algorithm for Cedille Core is presented as ``almost'' algorithmic typing rules (more on this shortly),
in Figure~\ref{fig:rules}.  The rules are almost algorithmic because we must understand a few conventions for applying the rules:

\begin{itemize}
\item The rules are applied bottom-up, and a judgment $\Gamma\vdash t : t'$ represents a call to compute
  a type $t'$ given a context $\Gamma$ and a term $t$.
\item Some of the premises of the rules state that a computed type should have a certain form (e.g., be a $\Pi$-type).  It may happen, though, that the computed type $\beta$-reduces to a $\Pi$-type but is not literally one.  To handle this case, one should head-normalize the computed types for premises that require a specific form.
\item Where two premises use the same meta-variable, one must check definitional equality of the terms in question (for example, the domain-part of a $\Pi$-type and the type of an argument, in the application typing rule).
  The definitional equality relation, which we can denote $\Gamma \vdash t =_{\beta\eta} t'$,
  is the usual relation of $\beta\eta$-equivalence of
terms in pure untyped lambda calculus, extended congruentially for the
typing constructs $\forall$, $\Pi$, $\iota$, and $\simeq$, and also
extended to make use of let-definitions $x = t : t'$ contained in
$\Gamma$ (by replacing $x$ with $t$ when checking definitional
equality).  This same mechanism can be used for global definitions, as
well.
\item Note that the formation rules for $\forall$- and $\Pi$-abstractions use
  a premise $\textit{Var}(x,s)$ to check
  that the variable $x$ is a legal form of variable given the kind $s$.  This
  judgment is defined by these rules:
  \[
  \begin{array}{lll}
    \infer{\textit{Var}(u,\star)}{\ } & \ \ \ & \infer{\textit{Var}(X,\Box)}{\ }
  \end{array}
  \]
  
\end{itemize}

\begin{figure}
  \[
  \begin{array}{lll}
    \infer{\Gamma\vdash \star : \Box}{\ } &
    \infer{\Gamma\vdash\abs{\Pi}{x}{t}{t'} : s'}{\Gamma \vdash t : s \quad \Gamma,x:t\vdash t' : s' \quad \textit{Var}(x,s)} &
    \infer{\Gamma\vdash\abs{\forall}{x}{t}{t'} : \star}{\Gamma \vdash t : s \quad \Gamma,x:t\vdash t' : \star \quad \textit{Var}(x,s)} 
    \\ \\
    \infer{\Gamma, x : t \vdash x : t}{\ } &
    \infer{\Gamma\vdash t\ t' : [t'/x]t_2}{\Gamma\vdash t : \abs{\Pi}{x}{t_1}{t_2} & \Gamma\vdash t : t_1} &
    \infer{\Gamma\vdash\abs{\iota}{u}{t}{t'} : \star}{\Gamma \vdash t : \star \quad \Gamma,x:t\vdash t' : \star} 
    \\ \\
    \infer{\Gamma\vdash t.2 : [t/x]t_2}{\Gamma\vdash t : \abs{\iota}{x}{t_1}{t_2}} &    
    \infer{\Gamma\vdash\abs{\lambda}{x}{t}{t'} : \abs{\Pi}{x}{t}{t''}}{\Gamma,x:t\vdash t':t'' \quad \Gamma \vdash \abs{\Pi}{x}{t}{t''} : s}  &
    \infer{\Gamma\vdash \varsigma\ t : \{ t_2 \simeq t_1\}}{\Gamma\vdash t : \{t_1 \simeq t_2\}}

\\ \\
    \infer{\Gamma\vdash t.1 : t_1}{\Gamma\vdash t : \abs{\iota}{x}{t_1}{t_2}} &
    \infer{\Gamma\vdash t\ \mhyph t' : [t'/x]t_2}{\Gamma\vdash t : \abs{\forall}{x}{t_1}{t_2} & \Gamma\vdash t : t_1} &
    \infer{\Gamma\vdash \rho\ t\ @\ x.t'\ \mhyph\ t'' : [t_1/x]t'}{\Gamma \vdash t : \{ t_1 \simeq t_2 \} \quad \Gamma\vdash t'' : [t_2/x]t'}
    \\ \\
    \infer{\Gamma\vdash \beta\ t'\{t\} : \{ t' \simeq t'\}}{\Gamma\vdash \{ t' \simeq t' \} : \star } &
    \infer{\Gamma\vdash \{ p \simeq p'\} : \star}{\textit{FV}(p\ p')\subseteq \textit{dom}(\Gamma)} &
     \infer{\Gamma\vdash [ k = t_1 : \Box ]\ \mhyph\ t_2 : t''}{\Gamma\vdash t_1 : \Box \quad
                                                               \Gamma, k = t_1 : t' \vdash t_2 : t''}  
    \\ \\
    \multicolumn{2}{l}{
    \infer{\Gamma\vdash [t,t'\ @\ x.t_2] : \abs{\iota}{x}{t_1}{t_2}}{\Gamma\vdash t : t_1 \quad \Gamma\vdash t' : [t/x]t_2 \quad \Gamma \vdash \abs{\iota}{x}{t_1}{t_2} : \star \quad |t| = |t'|}} &  
  \infer{\Gamma\vdash \phi\ t\ \mhyph\ t_1\ \{t_2\} : t'}{\Gamma \vdash t : \{t_1 \simeq t_2\} \quad \Gamma\vdash t_1 : t'}  
    \\ \\

    \multicolumn{2}{l}{
    \infer{\Gamma\vdash [ x = t_1 ]\ \mhyph\ t_2 : t''}{\Gamma\vdash t_1 : t' \quad \Gamma \vdash t' : s \quad \textit{Var}(x,s) \quad
      \Gamma, x = t_1 : t' \vdash t_2 : t''} } &
   \ 
    \\ \\
    \multicolumn{2}{l}{
      \infer{\Gamma\vdash\abs{\Lambda}{x}{t}{t'} : \abs{\forall}{x}{t}{t''}}{\Gamma,x:t\vdash t':t'' \quad x\not\in\textit{FV}(|t'|)\quad \Gamma \vdash \abs{\forall}{x}{t}{t''} : s}}&
    \infer{\Gamma\vdash \delta\ T\ t : T}{\Gamma\vdash t : \{ \absu{\lambda}{x}{\absu{\lambda}{y}{x}} \simeq \absu{\lambda}{x}{\absu{\lambda}{y}{y}}\} } 

  \end{array}
  \]
  \caption{Type-checking rules for Cedille Core}
  \label{fig:rules}
\end{figure}

Note that $\rho$ rewrites from the right-hand side of an equation to
the left-hand side, for consistency when synthesizing a type for a
$\rho$-term in full Cedille.

\end{document}